\documentclass[a4paper,fleqn,usenatbib,useAMS]{mnras}

\usepackage{graphicx}  
\usepackage{amsmath}  
\usepackage{amssymb}  
\usepackage{multicol}        
\usepackage{bm}    
\usepackage{pdflscape}  
\usepackage{multirow}
\usepackage{xcolor}
\usepackage{hyperref}
\usepackage{booktabs}
\usepackage{subfigure}
\usepackage{academicons}
\usepackage{orcidlink}
\newcommand{\orcid}[1]{\href{https://orcid.org/#1}{\textcolor[HTML]{A6CE39}{\aiOrcid}}}
\newcommand{\colibre}{\textsc{COLIBRE} }
\newcommand{\swift}{\textsc{SWIFT} }
\newcommand{\eagle}{\textsc{EAGLE} }
\newcommand{\logg}{\log_{10}}

\newcommand{\added}[1]{{{}#1}}

\newcommand{\kms}{\,km\,s$^{-1}$} 
\def\specialname[#1]{\textbf{\textsc{#1}}}

\usepackage[T1]{fontenc}
\usepackage{ae,aecompl}

\usepackage{newtxtext,newtxmath}

\title[The concentration dependence of SMHMR]{
  Gravitational potential drives the concentration dependence of the
  stellar mass-halo mass relation
}
\author[Wang et al.]{
Kai Wang,$^{1,2}$\thanks{wkcosmology@gmail.com}\orcidlink{0000-0002-3775-0484}
Joop Schaye,$^{3}$\orcidlink{0000-0002-0668-5560}
Alejandro Ben\'{i}tez-Llambay, $^{4}$\orcidlink{0000-0001-8261-2796}
Evgenii Chaikin, $^3$\orcidlink{0000-0003-2047-3684}
\newauthor
Carlos S. Frenk, $^{1}$\orcidlink{0000-0002-2338-716X}
Filip Hu\v{s}ko, $^{3}$\orcidlink{0000-0002-1510-1731}
Robert J. McGibbon, $^3$\orcidlink{0000-0003-0651-0776}
Sylvia Ploeckinger, $^5$\orcidlink{0000-0002-1965-1650}
\newauthor
Alexander J. Richings, $^{6, 7}$\orcidlink{0000-0003-0502-9235}
Matthieu Schaller, $^{8,3}$\orcidlink{0000-0002-2395-4902}
and James W. Trayford $^9$\orcidlink{0000-0003-1530-1634}
\newauthor
\\
$^{1}$Institute for Computational Cosmology, Department of Physics, Durham University, South Road, Durham, DH1 3LE, UK\\
$^{2}$Centre for Extragalactic Astronomy, Department of Physics, Durham University, South Road, Durham DH1 3LE, UK\\
$^{3}$Leiden Observatory, Leiden University, PO Box 9513, NL-2300 RA Leiden, the Netherlands\\
$^4$Dipartimento di Fisica G. Occhialini, Università degli Studi di Milano Bicocca, Piazza della Scienza, 3 I-20126 Milano MI, Italy\\
$^5$Department of Astrophysics, University of Vienna, Türkenschanzstrasse 17, 1180 Vienna, Austria\\
$^6$Centre for Data Science, Artificial Intelligence and Modelling, University of Hull, Cottingham Road, Hull, HU6 7RX, UK \\
$^7$E. A. Milne Centre for Astrophysics, University of Hull, Cottingham Road, Hull, HU6 7RX, UK\\
$^8$Lorentz Institute for Theoretical Physics, Leiden University, PO Box 9506, 2300 RA Leiden, the Netherlands\\
$^9$Institute of Cosmology and Gravitation, University of Portsmouth, Dennis Sciama Building, Burnaby Road, Portsmouth PO1 3FX, UK
}

\date{Last updated 2020 May 22; in original form 2018 September 5}

\pubyear{2020}

\begin{document}
\label{firstpage}
\pagerange{\pageref{firstpage}--\pageref{lastpage}}
\maketitle


\begin{abstract}
  We investigate the origin of the scatter in the stellar mass-halo mass (SMHM) relation using the \colibre cosmological hydrodynamical simulations.
  At fixed halo mass, we find a clear positive correlation between stellar mass and halo concentration, particularly in low-mass haloes between $10^{11}$ and $10^{12}\,\rm M_\odot$, where all halo properties are computed from the corresponding dark-matter-only simulation.
  Two scenarios have been proposed to explain this trend: the earlier formation of higher-concentration haloes allows more time for star formation, or the deeper gravitational potential wells of higher-concentration haloes enhance baryon retention.
  To distinguish between them, we examine correlations between halo concentration, stellar mass, stellar age, and stellar metallicity.
  While, at fixed halo mass, halo concentration correlates with stellar age, stellar age itself shows only a weak correlation with stellar mass, indicating that early formation alone cannot account for the concentration-dependence in the scatter of the SMHM relation.
  In contrast, both stellar metallicity and halo concentration exhibit  correlations with stellar mass.
  The connection between halo concentration and stellar metallicity persists even when simultaneously controlling for both halo mass and stellar mass.
  These results support the scenario in which the deeper gravitational potentials in higher-concentration haloes suppress feedback-driven outflows, thereby enhancing both baryon and metal retention.
\end{abstract}





\section{Introduction}%
\label{sec:introduction}

The connection between galaxies and their host dark matter haloes is a cornerstone of modern galaxy formation theory.
In the $\Lambda$CDM cosmological framework, galaxies form through the condensation of baryons within gravitational potential wells of dark matter haloes \citep{whiteGalaxyFormationHierarchical1991}, whose assembly is governed by hierarchical growth.
A central empirical relation that encodes this connection is the stellar mass-halo mass (SMHM) relation, which captures the efficiency with which haloes convert baryons into stars \citep[e.g.][]{ekeGalaxyGroupsTwodegree2004, yangConstrainingGalaxyFormation2003, mosterConstraintsRelationshipStellar2010, wechslerConnectionGalaxiesTheir2018, behrooziUNIVERSEMACHINECorrelationGalaxy2019}.
While the SMHM relation is relatively tight on average, significant intrinsic scatter persists at fixed halo mass in both observations \citep[e.g.][]{yangGALAXYGROUPSSDSS2009, moreSatelliteKinematicsNew2009, zuMappingStellarContent2015, kravtsovStellarMassHalo2018, scholz-diazDarkSideGalaxy2022, scholz-diazBaryonicPropertiesNearby2024} and simulations \citep[e.g.][]{guGuMengHIERARCHICALGALAXYGROWTH2016, tojeiroGalaxyMassAssembly2017, mattheeOriginScatterStellar2017, artaleImpactAssemblyBias2018, correaDependenceGalaxyStellartohalo2020, lyuHalosGalaxiesVII2023, peiEffectivenessHaloGalaxy2024}.
Understanding the origin of this scatter is crucial for understanding the connection between halo assembly and galaxy formation \citep[e.g.][]{mandelbaumGalaxyHaloMasses2006, wangScatterRelationStellar2013, cuiOriginGalaxyColour2021, zuDoesConcentrationDrive2021, wangLateformedHaloesPrefer2023, zhaoHalosGalaxies2025, wangTestingGalaxyFormation2025}, and for developing empirical models for galaxy formation \citep[e.g.][]{reddickConnectionGalaxiesDark2013, zehaviImpactAssemblyBias2018, zehaviProspectUsingMaximum2019}.

Previous simulation-based studies have highlighted the role of secondary halo properties—such as concentration, formation time, spin, and environment—in shaping the galaxy-halo connection beyond halo mass \citep[e.g.][]{mattheeOriginScatterStellar2017, wechslerConnectionGalaxiesTheir2018, boseRevealingGalaxyHalo2019, kulierEvolutionBaryonFraction2019, zehaviProspectUsingMaximum2019, martizziBaryonsCosmicWeb2020, montero-dortaManifestationSecondaryBias2020, bradshawPhysicalCorrelationsScatter2020}.
Among these, halo concentration, or equivalently the maximum circular velocity, which can be computed directly from halo mass and concentration, has emerged as a relatively strong correlate: at fixed halo mass, haloes with higher concentration tend to host central galaxies with larger stellar masses.
This trend is found in empirical models \citep[e.g.][]{bradshawPhysicalCorrelationsScatter2020}, semi-analytical models \citep[e.g.][]{tojeiroGalaxyMassAssembly2017, zehaviProspectUsingMaximum2019, peiEffectivenessHaloGalaxy2024}, hydrodynamical simulations \citep[e.g.][]{mattheeOriginScatterStellar2017, artaleImpactAssemblyBias2018, boseRevealingGalaxyHalo2019, martizziBaryonsCosmicWeb2020, peiEffectivenessHaloGalaxy2024}, \added{and observations \citep[e.g.][]{mancerapinaImpactGasDisc2022}}.
However, the physical origin of this dependence remains uncertain.

\citet{mattheeOriginScatterStellar2017} proposed two broad, physically motivated hypotheses to explain the concentration dependence of stellar mass at fixed halo mass.
The first links concentration to halo formation history: haloes that assemble earlier tend to be more concentrated \citep[e.g.][]{navarroUniversalDensityProfile1997, wechslerConcentrationsDarkHalos2002, zhaoACCURATEUNIVERSALMODELS2009, ludlowMassconcentrationredshiftRelationCold2014, correaAccretionHistoryDark2015}, and may thus have had more time to form stars.
In this scenario, stellar mass differences at fixed halo mass arise primarily from temporal effects, with earlier-forming haloes building more massive galaxies due to their more extended star formation histories.
The second hypothesis emphasizes gravitational potential depth: at fixed halo mass, more concentrated haloes have deeper gravitational potentials, which can reduce the ability of stellar feedback to eject gas.
This mechanism leads to more efficient gas retention and thus higher stellar mass.

Disentangling these two explanations is challenging.
While halo formation time and gravitational potential depth are correlated, they are not completely degenerate, and their respective imprints on galaxy properties may differ in subtle but measurable ways.
For example, one might expect the formation-time-driven scenario to imprint differences in stellar ages, whereas feedback-regulated growth may more directly influence stellar metallicity through differences in gas retention and recycling \citep[e.g.][]{pengHaloesGalaxiesDynamics2014, wangmetal}.

Here we investigate the origin of the concentration dependence of the SMHM relation using the new \colibre cosmological hydrodynamical simulation suite \citep{schayecolibre, chaikin2025a}.
We focus on central galaxies at $z=0$, and examine how galaxy stellar mass correlates with halo concentration, stellar age, and stellar metallicity at fixed halo mass.
By comparing the relative strength of these correlations, we aim to assess whether the primary driver of the SMHM scatter is the difference in the halo formation time or the gravitational potential depth.
This approach provides a physically motivated diagnostic of the two proposed mechanisms and offers a clear picture of how galaxy formation efficiency is modulated by halo structure.

The remainder of this paper is organized as follows.
\S\,\ref{sec:data} describes the \colibre simulations, the sample selection, and the physical quantities used in our analysis.
In \S\,\ref{sec:results}, we examine the correlation between stellar mass and halo concentration and its connection to stellar age and stellar metallicity, from which we can interpret these trends in the context of the proposed scenarios. \S\,\ref{sec:summary} presents the summary of our conclusions.
We assume the same $\Lambda$CDM cosmology as \colibre, which is based on the DES Y3 results \citep{abbottDarkEnergySurvey2022}, where $h=0.681$, $\Omega_{\rm m} = 0.306$, $\Omega_{\rm b}=0.0486$, $\Omega_{\Lambda} = 0.693922$.

\section{Simulations} \label{sec:data}

This study is based on the \colibre simulation suite, which consists of a set of hydrodynamical simulations and their corresponding dark-matter-only (DMO) counterparts with different box sizes and resolutions \citep{schayecolibre, chaikin2025a}.
The \colibre simulations were run with the \swift code \citep{schallerSWIFTModernHighlyparallel2024} with significant improvements to all subgrid prescriptions.
The most important new features include allowing radiative cooling below $10^4\,$K \citep{ploeckingerHybridchimesModelRadiative2025}, star formation with a gravitational instability criterion \citep{nobelsTestsSubgridModels2023}, improved chemical enrichment \citep{Correa2025}, simulating the growth and composition of dust grains with coupling to self-shielding and molecules \citep{trayfordModellingEvolutionInfluence2025}, modelling pre-supernova stellar feedback \citep{benitezllambay2025}, and improving the sampling of supernova and AGN feedback \citep{chaikinThermalkineticSubgridModel2023, schayecolibre}.
Most of the simulations have two flavours of AGN feedback: one thermally-driven AGN feedback \citep{boothCosmologicalSimulationsGrowth2009}, and one hybrid thermally-driven/kinetic jet-driven AGN feedback \citep{husko2025}.
Spurious transfer of energy from dark matter to baryons is suppressed by using four times more dark matter particles than baryonic particles \citep{ludlowSpuriousHeatingStellar2023}.

All \colibre simulations are calibrated to match the observed $z\approx 0$ galaxy stellar mass function, stellar mass-size relation, and stellar mass-black hole mass relation \citep{chaikin2025a}.
\colibre reproduces other properties and scaling relations that are not used in calibration, including the stellar and gas-phase metallicity relations, cosmic star formation history, galaxy quiescent fraction, atomic and molecular gas mass, X-ray luminosity from the circumgalactic medium \citep{schayecolibre}, and the evolution of the galaxy stellar mass function from $z>10$ to $z=0$ \citep{chaikin2025b}.

Here we focus on the \texttt{L200m6} simulation with thermally-driven AGN feedback, which has a cubic length of 200 comoving Mpc.
The masses of dark matter particles and baryonic particles are $2.42\times 10^6\,\rm M_\odot$ and $1.84\times 10^6\,\rm M_\odot$, respectively.
Haloes are identified using the Friends-of-Friends (FoF) algorithm, and subhaloes are identified and tracked using the HBT-HERONS method \citep{forouharmorenoAssessingSubhaloFinders2025}.
Halo and galaxy properties are calculated using the \texttt{SOAP} pipeline \citep{mcgibbonSOAPPythonPackage2025}.
Halo masses are defined as the total mass enclosed within a sphere whose mean density is 200 times the critical density of the universe.
Halo concentration, which characterizes the shape of the NFW density profile \citep{navarroUniversalDensityProfile1997}, is estimated from the first moment of the dark matter density profile \citep{wangEfficientRobustMethod2024}.
The stellar properties of galaxies are calculated using all bound stellar particles within 50 physical kpc of the most-bound particle of each subhalo.
Both stellar age and stellar metallicity are mass-weighted averages over all stellar particles for each galaxy.

To characterize the dark matter halo properties independently of baryonic effects, such as halo contraction \citep[e.g.][]{blumenthalContractionDarkMatter1986, schallerBaryonEffectsInternal2015}, we follow \citet{mattheeOriginScatterStellar2017} and use the corresponding DMO simulation with the same initial conditions to compute the halo mass, $M_{\rm 200c}^{\rm DMO}$, and concentration, $c^{\rm DMO}$.
Subhaloes in the two simulations are matched by identifying the 50 most bound dark matter particles in each subhalo.
This matching yields a bijective success rate of over $98\%$ for central subhaloes with masses above $10^{11}\,\rm M_\odot$, ensuring a robust mapping between halo properties in the DMO and hydrodynamical simulations\footnote{
  The {\tt L200m6} DMO run experienced a disk failure at $z\approx 0.22$ and was restarted from the most recent snapshot.
  This restart introduced minor discontinuities in the time integration of some particle trajectories, causing a few massive haloes in the hydrodynamical run to be spuriously matched to low-mass haloes in the DMO run.
  To remove these artefacts, we exclude all matched haloes with mass differences exceeding 0.3 dex.
  After this filtering, the bijective matching success rate remains
  $\gtrsim 98\%$ for haloes with $10^{11-12}\,\rm M_\odot$, and we verified that this has no impact on the results presented in this paper.}.

In this work, we focus exclusively on central galaxies\footnote{
  As detailed in \citet{forouharmorenoAssessingSubhaloFinders2025}, all HBT-HERONS subhaloes are formed initially as centrals.
  During the subsequent hierarchical merging, the central subhalo is selected to be the one whose bound mass is at least 80\% of the most massive progenitor central subhalo, and, if multiple subhaloes survive this threshold, the one with the lowest specific orbital kinetic energy in the frame of the host FoF halo is the new central.} at $z=0$ residing in haloes with the corresponding DMO mass $M_{\rm 200c}^{\rm DMO}>10^{11}\,\rm M_\odot$.
This selection yields a sample of $\gtrsim 77,000$ haloes.
These haloes and their central galaxies are both well-resolved with $\gtrsim 50,000$ and $\gtrsim 100$ dark matter and stellar particles respectively.
\added{Correspondingly, central galaxies in haloes with $M^{\rm DMO}_{\rm 200c} \gtrsim 10^{11.5}\rm M_\odot$ and $10^{12}\rm M_\odot$ are resolved by $\gtrsim 1,000$ and $10,000$ stellar particles, respectively.
  These better-resolved subsamples yield consistent results, further supporting the robustness of our conclusions.
}

\section{Results}%
\label{sec:results}

\subsection{The stellar mass-halo mass relation and its dependence on halo
  concentration}
\label{sub:dependence_on_concentration} 

\begin{figure*}
  \begin{center}
    \includegraphics[width=0.9\linewidth]{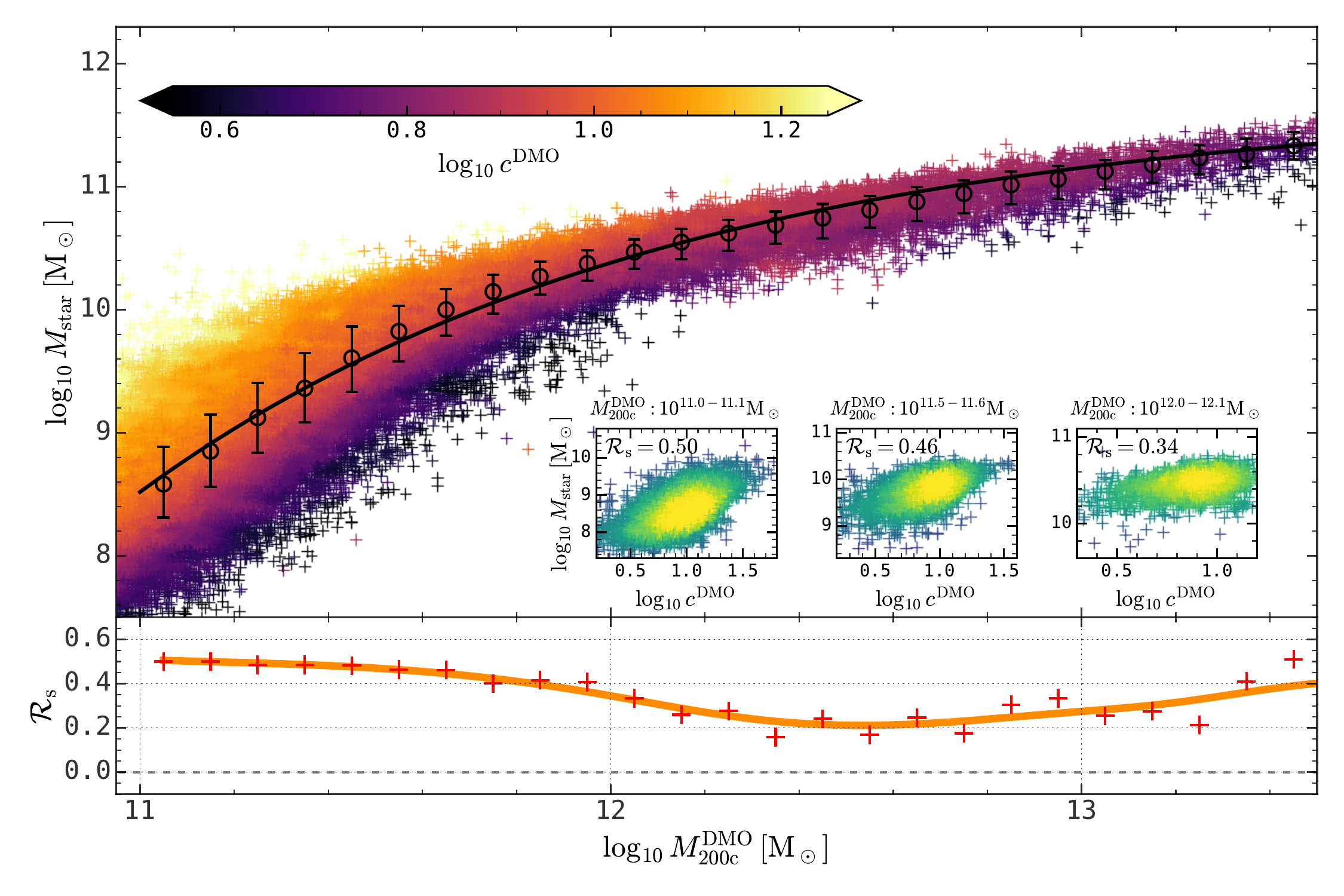}
  \end{center}
  \caption{
    The SMHM relation for central galaxies in the \colibre \texttt{L200m6} simulation, where colour encodes the logarithm of the halo concentration smoothed over 0.1-dex stellar mass and halo mass bins.
    Here both halo mass and halo concentration come from the corresponding DMO run.
    The error bars show the median and the $16^{\rm th}-84^{\rm th}$ stellar mass quantiles within 0.1-dex-width halo mass bins.
    The solid line shows the best-fitting result with the functional form in equation~(\ref{eq:fitting}).
    Three inset panels show the scatter plot, as well as Spearman's rank correlation coefficient, between stellar mass and halo concentration in three selected 0.1-dex-width halo mass bins, where the colour encodes the density of the scatter point distribution.
    The red crosses in the bottom panel show Spearman's rank correlation coefficients between stellar mass and halo concentration in 0.1-dex-width halo mass bins, where only bins with more than 30 data points are shown.
    The solid line is a smoothing B-spline line fitted to the data points to show the trend.
    There is a moderately strong correlation between stellar mass and concentration at fixed halo mass, particularly below $10^{12}\,\rm M_\odot$, and the strength declines above this mass.
  }
  \label{fig:shmr}
\end{figure*}

We begin our analysis by examining the relationship between stellar mass and halo mass for central galaxies, and its dependence on halo concentration.
Fig.~\ref{fig:shmr} shows the SMHM relation for central galaxies at $z=0$.
Following \citet{mattheeOriginScatterStellar2017}, halo masses and concentrations are taken from the matched haloes in the corresponding DMO simulation to better isolate the causal relationship, as the dark matter distribution in the hydrodynamical runs is influenced by baryonic processes \citep[e.g.][]{duffyImpactBaryonPhysics2010, schallerBaryonEffectsInternal2015}.
As expected, the median stellar mass increases monotonically across bins of increasing halo mass, though individual haloes exhibit noticeable scatter from the median relation, particularly in the low-mass regime.
We compute the median and 16th-84th percentile range of stellar mass in bins of 0.1 dex in halo mass, and fit the median relation with equal weighting in 0.1-dex halo mass bins from $10^{11}$ to $10^{13.5}\,\rm M_\odot$ using an empirical function \citep{mattheeOriginScatterStellar2017}:
\begin{equation}
  \logg \left(\frac{M_{\rm star}}{\rm M_\odot}\right) = 11.70 -
  \exp\left[-0.88\logg\left(\frac{M_{\rm 200c}^{\rm DMO}}{\rm M_\odot}\right) -
    10.85\right],
  \label{eq:fitting}
\end{equation}
which provides a good description of the overall trend across the full mass range considered.

Overlaid on the SMHM relation, we use colour to encode the concentration of the matched DMO halo for each galaxy.
There is a moderately strong secondary dependence of stellar mass on halo concentration: at fixed halo mass, central galaxies residing in higher-concentration haloes exhibit systematically higher stellar masses than their lower-concentration counterparts.
This concentration dependence is most pronounced in low-mass haloes.
To quantify this effect, the bottom panel of Fig.~\ref{fig:shmr} shows Spearman's rank correlation coefficient\footnote{Spearman's rank correlation coefficient quantifies the rank correlation between two variables and ranges from 0 for no correlation to +1/-1 for a strictly monotonic relation.}, $\mathcal R_{\rm s}$, between stellar mass and DMO halo concentration as a function of halo mass with a bin size of 0.1 dex.
We also present three scatter plots between halo concentration and stellar mass in three selected halo mass bins in the smaller inset panels of Fig.~\ref{fig:shmr} for a visual impression of the correlation strength.
We find a moderately strong and statistically significant correlation in low-mass haloes, with coefficients of $\mathcal R_{\rm s}\approx 0.50$ in the $M_{\rm 200c}^{\rm DMO} = 10^{11.0-11.1}\,\rm M_\odot$ bin and $\mathcal R_{\rm s}\approx 0.46$ in the $M_{\rm 200c}^{\rm DMO} =
  10^{11.5-11.6}\,\rm M_\odot$ bin.
The correlation becomes weaker at higher halo mass, with a value of $\mathcal R_{\rm s}\approx 0.15-0.35$ above $10^{12}\,\rm M_\odot$, though it remains positive.
These results are consistent with the findings of \citet{mattheeOriginScatterStellar2017} for the \eagle simulation \citep{schayeEAGLEProjectSimulating2015}.

\subsection{Disentangling the physical origin of the concentration dependence}
\label{sub:physical_origin} 

\begin{figure*}
  \begin{center}
    \includegraphics[width=1\linewidth]{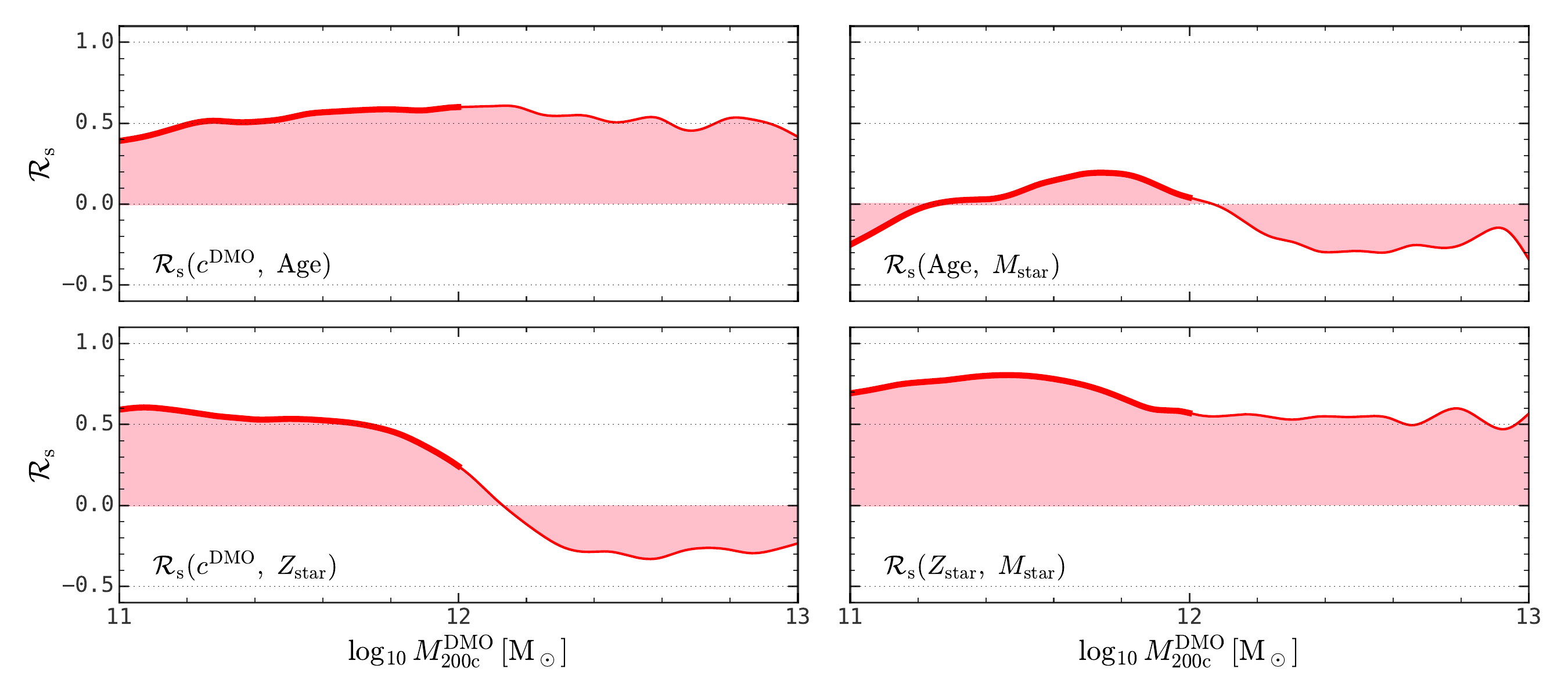}
  \end{center}
  \caption{
    Pairwise Spearman's rank correlation coefficients for relations between halo and central galaxy properties within 0.1-dex-width halo mass bins from $10^{11}$ to $10^{13}\,\rm M_\odot$.
    The four panels correspond to correlations between halo concentration and stellar age, stellar age and stellar mass, halo concentration and stellar metallicity, and stellar metallicity and stellar mass, respectively.
    Here both stellar age and stellar metallicity are weighted by the mass of stellar particles, and all halo related properties are from the matched haloes in the DMO run.
    In the halo mass range of $10^{11-12}\,\rm M_\odot$, there is at most a weak correlation between stellar age and stellar mass while all other correlations are of moderate strength, so the correlation between halo concentration and stellar mass is mediated by stellar metallicity rather than by stellar age.
  }
  \label{fig:correlation_age_zstar}
\end{figure*}

\begin{figure*}
  \begin{center}
    \includegraphics[width=0.9\linewidth]{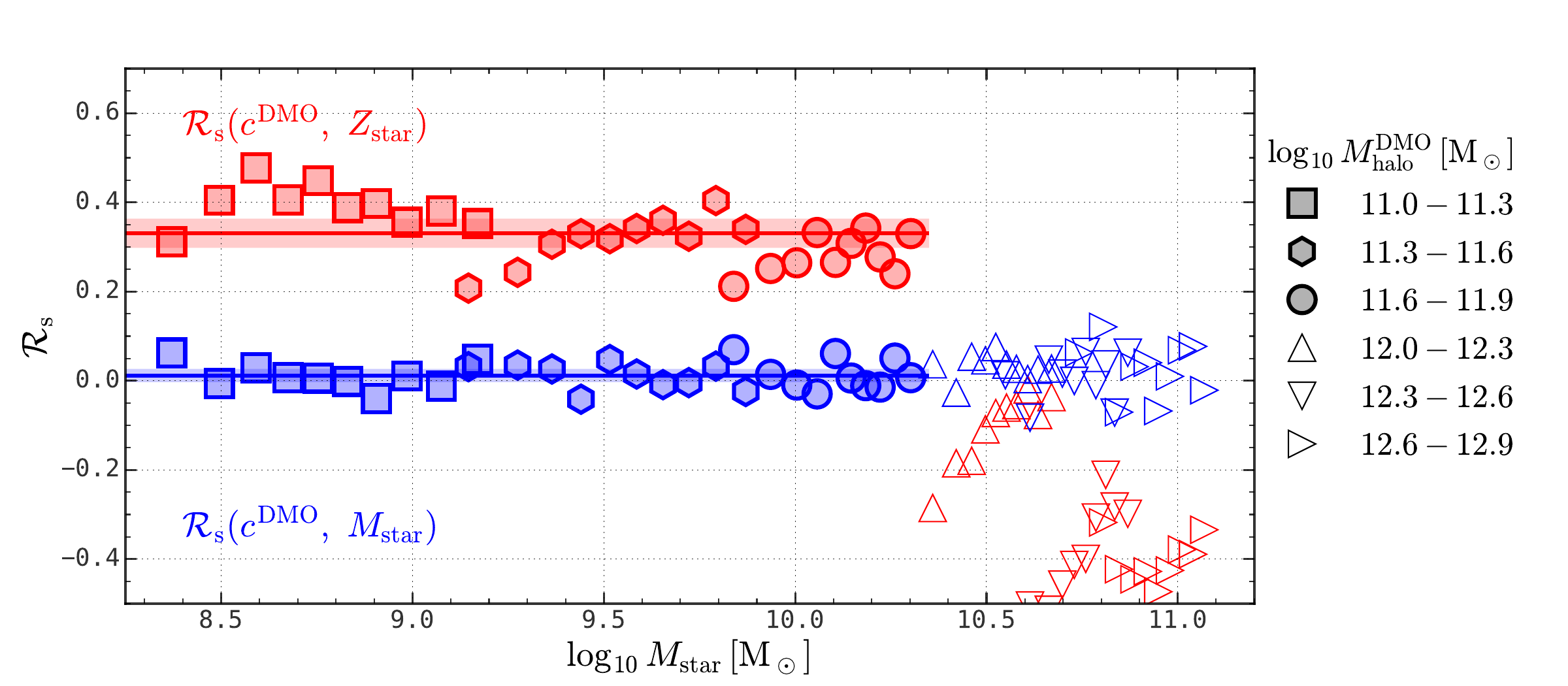}
  \end{center}
  \caption{
    Spearman's rank correlation coefficients between halo concentration and stellar metallicity (red symbols), and between halo concentration and stellar mass (blue symbols) at fixed stellar and halo mass.
    Results for haloes within the mass ranges $10^{11-12}\,\rm M_\odot$ and $10^{12-13}\,
      \rm M_\odot$ are shown with filled and open symbols, respectively.
    Stellar mass and halo mass are constrained within 0.1-dex and 0.3-dex bins, respectively.
    The two horizontal lines and the shaded regions are the mean values and standard deviations of the correlation coefficients for haloes of mass $10^{11-12}\,\rm M_\odot$.
    The correlation coefficient is only evaluated when more than 30 galaxies are present in the bin.
    All halo related quantities come from the matched haloes in the DMO run.
    As expected, there is no correlation between halo concentration and stellar mass when both halo mass and stellar mass are controlled.
    However, there is still a moderately strong correlation between concentration and stellar metallicity.
    The fact that this correlation is present at fixed stellar mass implies that it is due to the potential depth rather than the amount of metals produced.
  }
  \label{fig:correlation_fix_smhm}
\end{figure*}

The correlation between stellar mass and halo concentration at fixed halo mass shown in Fig.~\ref{fig:shmr} invites an investigation into its underlying cause. \citet{mattheeOriginScatterStellar2017} proposed two possible explanations for an analogous relation seen in EAGLE.
The first takes halo concentration as a proxy for halo formation time since early-forming haloes are typically more concentrated \citep{navarroUniversalDensityProfile1997, wechslerConcentrationsDarkHalos2002, zhaoACCURATEUNIVERSALMODELS2009, ludlowMassconcentrationredshiftRelationCold2014}, and thus have more time to convert gas into stars.
The second is based on the fact that higher-concentration haloes have deeper gravitational potentials, which reduces the effectiveness of stellar feedback and allows for more efficient retention of baryons, compared to lower-concentration haloes.

To distinguish these two scenarios, we analyze how halo concentration correlates with stellar age and stellar metallicity, and how both properties relate to stellar mass.
The motivation is that stellar age serves as a proxy for star formation history, so that if earlier halo assembly time results in higher stellar mass, we expect, at fixed halo mass, the correlation between halo concentration and stellar age with a strength similar to the correlation between concentration and stellar mass.
In contrast, stellar metallicity reflects the integrated effect of metal production and retention, so that in haloes with deeper gravitational potentials, i.e. higher concentrations, stellar feedback is less efficient at ejecting metal-enriched gas, leading to higher metallicities and higher stellar mass \citep[e.g.][]{
  pengHaloesGalaxiesDynamics2014, maRevisitingFundamentalMetallicity2024, jiaPotentialdrivenMetalCycling2025}.
Nevertheless, at fixed halo mass, a positive correlation between stellar mass and stellar metallicity could alternatively be attributed to more metal production during the star formation process.
However, we will shortly show that stellar metallicity correlates moderately with halo concentration even when we control for both halo mass and stellar mass.

We are mainly interested in the halo mass range from $10^{11}$ to $10^{12}\,\rm M_\odot$.
The lower limit is due to the mass resolution of the simulation, as we require at least $\approx 100$ stellar particles to minimally resolve the stellar population.
The upper limit is chosen to be $10^{12}\,\rm M_\odot$ for two reasons.
Firstly, the correlation between stellar mass and halo concentration above this halo mass becomes very weak \citep[see Fig.~\ref{fig:shmr} and also][]{mattheeOriginScatterStellar2017, peiEffectivenessHaloGalaxy2024}.
Secondly, AGN feedback starts to play a major role in the evolution of central galaxies above this halo mass \citep[e.g.][]{bowerDarkNemesisGalaxy2017}.
Previous studies found that the growth of super-massive black holes correlates with halo concentration, which is also attributed to the gravitational potential depth \citep{boothDarkMatterHaloes2010, boothUnderstandingEvolutionScaling2011}.
Meanwhile, AGN feedback can suppress star formation activity \citep[e.g.][]{bowerBreakingHierarchyGalaxy2006, dimatteoDirectCosmologicalSimulations2008} and expel the metal-rich gas from the galaxy center and reduce the global metallicity \citep[e.g.][]{derossiGalaxyMetallicityScaling2017, wangEnvironmentalDependenceMassMetallicity2023}.
The additional physical processes involved in massive haloes complicate the interpretation, so we will leave this regime to future work.

The correlations between halo concentration, stellar mass, stellar metallicity, and stellar age are summarized in Fig.~\ref{fig:correlation_age_zstar}, which shows pairwise Spearman's rank correlation coefficients between relevant properties measured in 0.1-dex halo mass bins across $10^{11-13}\,\rm M_\odot$.
For the low-mass regime from $10^{11}$ to $10^{12}\,\rm M_\odot$, the top two panels reveal a moderately strong positive correlation between halo concentration and stellar age that increases with halo mass from $\mathcal R_{\rm s}\approx 0.40$ to $\approx 0.60$, but only a weak correlation between stellar age and stellar mass, ranging from $\mathcal R_{\rm s}\approx -0.2$ to $\approx 0.2$, which is even negative below $10^{11.2}\,\rm M_\odot$ \citep[see][for consistent results in other simulations and semi-analytical models]{peiEffectivenessHaloGalaxy2024}.
This suggests that while galaxies in higher-concentration haloes formed earlier, this does not by itself lead to systematically higher stellar masses, falsifying the early-formation hypothesis.

The bottom two panels of Fig.~\ref{fig:correlation_age_zstar} show moderately strong and coherent trends involving stellar metallicity for halo masses between $10^{11}$ and $10^{12}\,\rm M_\odot$.
The rank correlation coefficient between halo concentration and stellar metallicity ranges from $\mathcal R_{\rm s}\approx 0.60$ to $\approx 0.30,$ and the stellar metallicity-stellar mass correlation remains consistently between $\mathcal R_{\rm s}\approx 0.70$ and $\approx 0.60$ across the halo mass range $10^{11}-10^{12}\,\rm M_\odot$.
These aligned trends suggest that the deeper gravitational potentials of higher-concentration haloes enhance metal retention and reduce the impact of feedback, enabling more efficient star formation and, consequently, higher stellar masses.
The behavior changes in haloes above $10^{12}\,\rm M_\odot$ due to the involvement of AGN feedback; we leave this for future work.

Although the results in Fig.~\ref{fig:correlation_age_zstar} are consistent with the potential-driven scenario, it remains possible that the observed correlation between halo concentration and stellar metallicity arises indirectly through their mutual dependence on stellar mass.
To address this possibility, Fig.~\ref{fig:correlation_fix_smhm} shows Spearman's rank correlation coefficient for the relation between stellar metallicity and halo concentration while simultaneously controlling for both stellar mass and halo mass by binning central galaxies in narrow intervals of stellar mass (0.1 dex) and halo mass (0.3 dex).
As a consistency check, we also measure the correlation between stellar mass and concentration within the same bins to verify that stellar mass is effectively held fixed.

Fig.~\ref{fig:correlation_fix_smhm} shows that the correlation between stellar mass and concentration disappears under these controls, confirming the successful mitigation of stellar mass effects.
In contrast, the concentration-stellar metallicity correlation remains moderately strong with a value of $\mathcal R_{\rm s}\approx 0.34$ for haloes below $10^{12}\,\rm M_\odot$.
This confirms that the stellar metallicity-concentration relation is not a byproduct of the correlation between stellar metallicity and stellar mass, and between halo concentration and stellar mass.
Instead, this indicates a direct connection between halo potential depth and metal retention efficiency.
These results thus reinforce the interpretation that deeper gravitational potentials in higher-concentration haloes reduce feedback-driven outflows, enhancing metal retention and leading to higher stellar masses.
Again, in haloes with masses above $10^{12}\,\rm M_\odot$, AGN feedback becomes more important than stellar feedback and changes the correlation between halo concentration and stellar metallicity.

\section{Discussion and summary}
\label{sec:summary}

In this work, we investigate the origin of the scatter in the SMHM relation using the \texttt{L200m6} \colibre cosmological hydrodynamical simulation.
Our analysis focuses on the role of halo concentration as a secondary parameter, and confirms the findings of \citet{mattheeOriginScatterStellar2017} that halo concentration is moderately correlated with stellar mass at fixed halo mass, particularly for haloes in the mass range $10^{11}-10^{12}\,\rm M_\odot$ (see Fig.~\ref{fig:shmr}).
We examine whether this correlation is primarily driven by earlier halo formation time or by deeper gravitational potential resulting in less efficient feedback.

We compare the pairwise correlations among halo concentration, stellar age, stellar metallicity, and stellar mass, all with halo mass fixed in 0.1-dex mass bins.
We find that, for haloes of mass between $10^{11}$ and $10^{12}\,\rm M_\odot$, although stellar age correlates moderately with halo concentration, it correlates only weakly with stellar mass.
In contrast, stellar metallicity exhibits moderately strong correlations with both halo concentration and stellar mass (see Fig.~\ref{fig:correlation_age_zstar}).
This suggests that the deeper potential wells of higher-concentration haloes enhance metal retention and star formation efficiency, and give rise to a correlation between halo concentration and galaxy stellar mass, rather than early formation time being responsible.

Since a higher stellar mass could directly cause a higher metallicity through the larger amount of metals produced, we further test our interpretation by computing the stellar metallicity-concentration correlation while simultaneously controlling for both stellar mass and halo mass.
The correlation remains significant with a Spearman's rank correlation coefficient of $\approx 0.34$ (see Fig.~\ref{fig:correlation_fix_smhm}).
This confirms that the concentration-stellar metallicity link is not mediated through stellar mass and supports a direct and important role of halo potential depth in shaping galaxy properties at fixed halo mass.

We verify the robustness of our results using the \colibre runs with hybrid AGN feedback \citep[see][]{husko2025}, as well as simulations at different resolutions \citep[see][]{schayecolibre} (see Appendix~\ref{sec:alternative_resolutions_and_agn_feedback}).
The {\tt m5} and {\tt m7} runs have average baryonic particle masses of $2.30\times 10^5\rm M_\odot$ and $1.47\times 10^7\rm M_\odot$, and the corresponding dark matter particle masses are $3.03\times 10^5\rm M_\odot$ and $1.94\times 10^7\rm M_\odot$, respectively.
The correlation trends obtained from all these runs are consistent with those in the {\tt L200m6} simulation, confirming that our conclusions are insensitive to the adopted AGN feedback model and to numerical resolution.

We recognise that stellar age and metallicity are correlated, as the stellar metallicity is shaped by the star formation history.
However, it is unlikely that, at fixed halo mass, the stellar mass-metallicity correlation is driven by stellar age, given that stellar age exhibits only a weak correlation with stellar mass (see Fig.~\ref{fig:correlation_age_zstar}).
Therefore, the age-metallicity relation does not compromise our main conclusion regarding the role of halo concentration and its effects on gravitational potential in shaping stellar metallicity and mass.
Our results therefore provide an appealing explanation for the physical origin of the scatter in the fundamental relation between central galaxy stellar masses and the masses of the haloes in which they reside.

We also find that haloes with masses above $10^{12}\,\rm M_\odot$ exhibit different behavior compared to lower-mass ones.
In this high-mass regime, AGN feedback plays a more dominant role than stellar feedback in regulating galaxy evolution.
A detailed investigation of this mass range is beyond the scope of the current work and will be explored in future studies.

Finally, our results offer promising avenues for observational validation, as stellar metallicity can be derived from spectroscopic observations and halo mass and concentration can be inferred through gravitational lensing and galaxy kinematics.
This makes the predicted connection between stellar metallicity and halo potential depth directly testable with current and upcoming observations.

\section*{DATA AVAILABILITY}

The data supporting the plots in this article are available on reasonable request to the corresponding author.
The \colibre simulation data will eventually be made publicly available, although we note that the data volume
(several petabytes) may prevent us from simply placing the raw data on a server.

\section*{Acknowledgements}

KW thanks Katherine Harborne, Tom Theuns, and Yangyao Chen for helpful discussions.
KW acknowledges support from the Science and Technologies Facilities Council (STFC) through grant ST/X001075/1.
FH acknowledges funding from the Netherlands Organization for Scientific Research (NWO) through research programme Athena 184.034.002.
ABL acknowledges support by the Italian Ministry for Universities (MUR) program “Dipartimenti di Eccellenza 2023-2027” within the Centro Bicocca di Cosmologia Quantitativa (BiCoQ), and support by UNIMIB’s Fondo Di Ateneo Quota Competitiva (project 2024-ATEQC-0050).
JT acknowledges support of a STFC Early Stage Research and Development grant (ST/X004651/1).
This project has received funding from the Netherlands Organization for Scientific Research (NWO) through research programme Athena 184.034.002.
CSF acknowledges support from European Research Council (ERC) Advanced Grant DMIDAS (GA 786910).

This work used the DiRAC@Durham facility managed by the Institute for Computational Cosmology on behalf of the STFC DiRAC HPC Facility (www.dirac.ac.uk).
The equipment was funded by BEIS capital funding via STFC capital grants ST/K00042X/1, ST/P002293/1, ST/R002371/1 and ST/S002502/1, Durham University and STFC operations grant ST/R000832/1.
DiRAC is part of the National e-Infrastructure.

The computation in this work is supported by swiftsimio \citep{2020JOSS....5.2430B}, SWIFTGalaxy \citep{2025JOSS...10.9278O}, and \specialname[HIPP] \citep{hipp}.
This research made use of NASA’s Astrophysics Data System for bibliographic information.

\bibliographystyle{mnras}
\bibliography{bibtex.bib}

\appendix

\section{Alternative resolutions and AGN feedback}
\label{sec:alternative_resolutions_and_agn_feedback}

\begin{figure*}
  \begin{center}
    \includegraphics[width=0.9\linewidth]{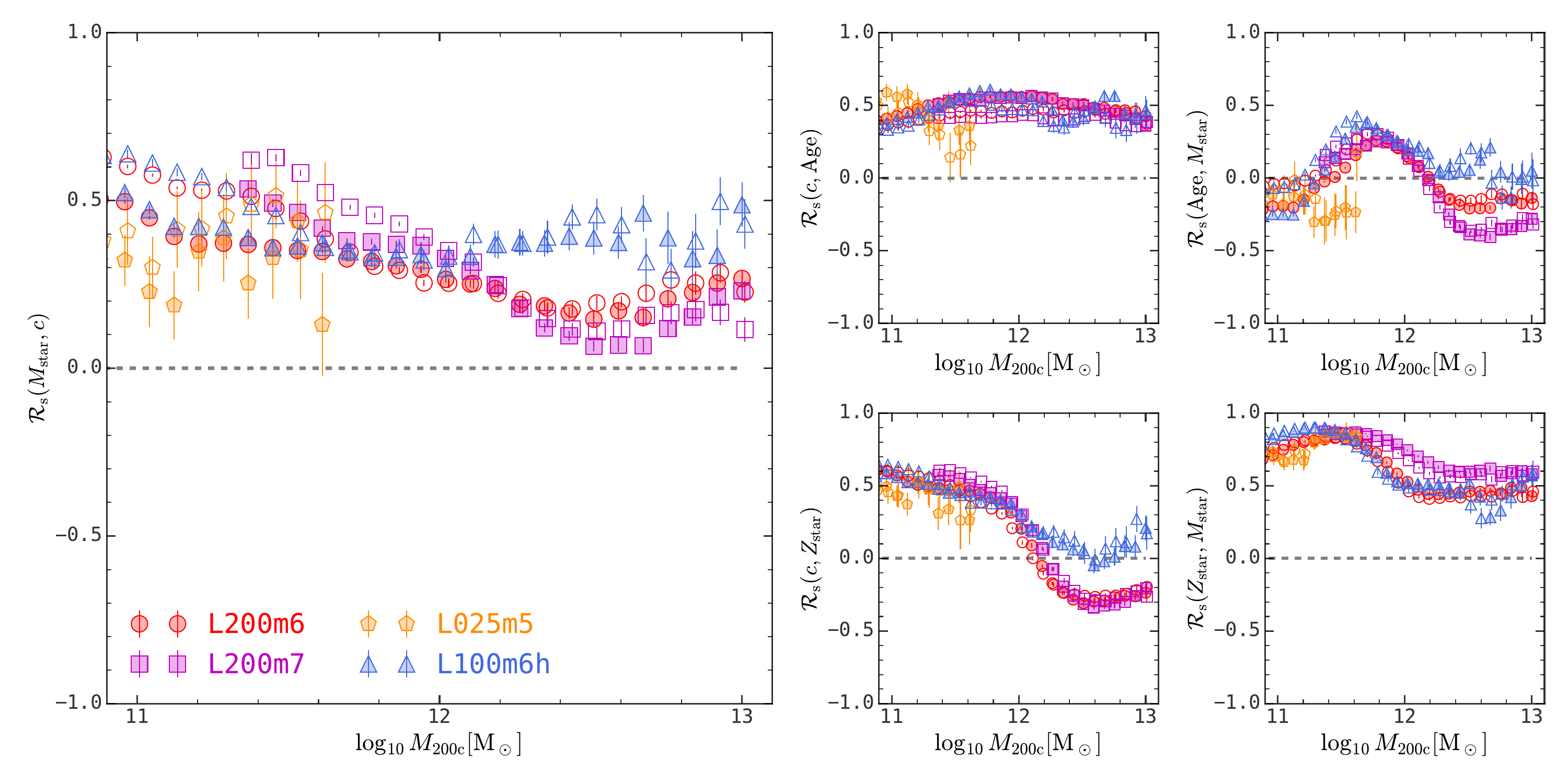}
  \end{center}
  \caption{
    \added{
      Pairwise rank correlation coefficient between stellar mass and halo concentration (left big panel) within 0.3-dex-width halo mass bins from $10^{11}$ to $10^{13}\,\rm M_\odot$.
      The four small panels correspond to correlations between halo concentration and stellar age, stellar age and stellar mass, halo concentration and stellar metallicity, and stellar metallicity and stellar mass, respectively.
      Results are shown for three different simulation resolutions (\texttt{L200m6}, \texttt{L200m7}, and \texttt{L025m5}), and for the hybrid AGN feedback (\texttt{L100m6h}).
      Filled symbols use halo mass and concentration from the corresponding DMO simulation, and open symbols use halo mass and concentration from the hydrodynamical simulation.
      Error bars show the standard deviation from 100 bootstrap realisations.
      Results for \texttt{L200m7} are shown only for $M_{\rm halo}^{\rm DMO} > 10^{11.5}\rm M_\odot$, where central galaxies are resolved with $\gtrsim 100$ stellar particles.
    }
  }
  \label{fig:correlation_comparison}
\end{figure*}

\begin{table}
  \caption{\colibre hydrodynamical simulations. The columns list the simulation identifier; the comoving box side length, $L$; the initial mean baryonic particle mass, $m_{\rm g}$; the mean CDM particle mass, $m_{\rm CDM}$; the AGN feedback prescription. We refer readers to \citet{schayecolibre} for details.}\label{tab:parameter}
  \begin{center}
    \begin{tabular}[c]{c|c|c|c|c}
      \hline
      Identifier       & $L$/cMpc & $m_g/\rm M_\odot$ & $m_{\rm CDM}/\rm M_\odot$ & AGN feedback \\
      \hline
      \texttt{L200m6}  & 200      & $1.84\times 10^6$ & $2.42\times 10^6$         & Fiducial     \\
      \texttt{L200m7}  & 200      & $1.47\times 10^7$ & $1.94\times 10^7$         & Fiducial     \\
      \texttt{L025m5}  & 50       & $2.30\times 10^5$ & $3.03\times 10^5$         & Fiducial     \\
      \texttt{L100m6h} & 100      & $1.84\times 10^6$ & $2.42\times 10^6$         & Hybrid       \\
      \hline
    \end{tabular}
  \end{center}
\end{table}

\added{

  Fig.~\ref{fig:correlation_comparison} compares the pairwise Spearman's rank correlation coefficients between halo and galaxy properties in high- and low-resolution simulations, \texttt{L025m5} and \texttt{L200m7}, respectively, and in the hybrid AGN simulation, \texttt{L100m6h}, as a function of $M^{\rm DMO}_{\rm 200c}$.
  For \texttt{L200m7} we show results only above $10^{11.5}\,\rm M_\odot$, where central galaxies are resolved with $\gtrsim 100$ stellar particles.
  Table~\ref{tab:parameter} lists the simulation parameters.

  Across all simulations, the correlation between $M_{\rm star}$ and the DMO concentration, $c^{\rm DMO}$, stays near $\mathcal R_{\rm s}\simeq 0.4-0.5$ over $M^{\rm DMO}_{\rm 200c} \sim 10^{11}-10^{12}\,\rm M_\odot$. The \texttt{L025m5} results show larger fluctuations due to the smaller number of haloes in this volume.
  Halo concentration also correlates with both stellar age and stellar metallicity, with $\mathcal R_{\rm s}\approx 0.5$ in all suites.
  By contrast, the age-mass correlation is weak, and becomes slightly negative around $M^{\rm DMO}_{\rm 200c} \sim 10^{11}\,\rm M_\odot$.
  The metallicity-mass correlation remains strong throughout, with $\mathcal R_{\rm s}\simeq 0.8$, consistent with the main text.

  We also present the results using halo mass and concentration measured in the hydrodynamical runs (open symbols), which are more directly comparable to observational inferences \citep[e.g.][]{mancerapinaImpactGasDisc2022}. This choice strengthens the correlation between halo concentration and stellar mass from $\simeq 0.4-0.5$ to $\simeq 0.5-0.6$, mainly because baryonic contraction increase the central mass density \citep[e.g.][]{schallerBaryonEffectsInternal2015}.

  The suites differ in resolution and AGN feedback implementation and are independently recalibrated to reproduce the $z=0$ stellar mass function and mass-size relation \citep{chaikin2025a, husko2025, schayecolibre}.
  Note that because the calibration is imperfect and limited to these observables, differences between simulations with different resolutions may not be directly caused by the differences in resolution.
  All simulations yield consistent rank correlation coefficients.
  The trends we found therefore do not depend sensitively on numerical resolution or on the details of the AGN prescription.

}

\bsp  
\label{lastpage}
\end{document}